\documentclass[prl,aps,twocolumn,showpacs,floatfix]{revtex4}
\usepackage{amsfonts}

\usepackage{amssymb}
\usepackage{hyperref}
\usepackage{graphicx}
\usepackage{dcolumn}
\usepackage{bm,amsmath,verbatim}
\usepackage{mathrsfs}
\usepackage{color}

\def\be{\begin{equation}}
\def\ee{\end{equation}}
\def\bea{\begin{eqnarray}}
\def\eea{\end{eqnarray}}
\def\bse{\begin{subequations}}
\def\ese{\end{subequations}}

\def\be{\begin{eqnarray}}
\def\ee{\end{eqnarray}}

\begin{document}

\title{Structured Weyl Points in Spin-Orbit-Coupled Fermionic Superfluids}
\author{Yong Xu}
\author{Fan Zhang}
\author{Chuanwei Zhang}
\email{chuanwei.zhang@utdallas.edu}

\begin{abstract}
We demonstrate that a Weyl point, widely examined in 3D Weyl semimetals and
superfluids, can develop a pair of non-degenerate gapless spheres. Such a
\emph{bouquet of two spheres} is characterized by \emph{three distinct}
topological invariants of manifolds with full energy gaps, i.e., the Chern
number of a 0D point inside one developed sphere, the winding number of a 1D
loop around the original Weyl point, and the Chern number of a 2D surface
enclosing the whole bouquet. We show that such structured Weyl points can be
realized in the superfluid quasiparticle spectrum of a 3D degenerate Fermi
gas subject to spin-orbit couplings and Zeeman fields, which supports
Fulde-Ferrell superfluids as the ground state.
\end{abstract}

\affiliation{Department of Physics, University of Texas at Dallas, Richardson, Texas
75080, USA}
\maketitle

Weyl fermions~\cite{Weyl} were initially conceived to describe neutrinos in
particle physics. Although neutrinos may have masses, the Standard Model of
particle physics permits the existence of such chiral fermions in quarks and
leptons~\cite{volovik}. Recently, Weyl fermions have been widely examined in
a class of solid-state materials dubbed Weyl semimetals~\cite%
{Murakami2007NJP,Wan2011prb,Yuanming2011PRB,Burkov2011PRL,ZhongFang2011prl,Aji2012prb,Burkov2012prb,Bernevig2012prl, LingLu2013NP,Xiaoliang2013,Vafek2014,Dai2014,Hasan2015}%
. Remarkably, these semimetals can be described by Weyl Hamiltonian near
their unusual Weyl points, where two linearly dispersed bands cross.
Impressively, such Weyl points have been experimentally observed in a
photonic crystal~\cite{Lu2015} and a Weyl semimetal TaAs~\cite{Xu2015,Lv2015}%
. A Weyl point can also be regarded as a monopole in 3D momentum space that
exhibits an integer topological charge, i.e., the quantized first Chern
number of a surface enclosing the singularity. Weyl points were also
suggested to exist in the quasiparticle spectrum of superfluid $^{3}\mathrm{%
He\;A}$ phase~\cite{volovik}. In contrast to traditional fully gapped
superfluids, the Weyl superfluids bear doubly degenerate nodes pinned to
zero energy, around which the quasiparticle energies disperse linearly in
all directions. Most recently, the existence of such Weyl nodes has also
been generalized to various cold-atom superfluids and solid-state
superconductors~\cite{Gong2011prl,Balents2012PRB,Tumanta2012PRB,
Das2013PRB,Sumanta2013PRA,Yong2014PRL,Yang2014PRL,Balicas2013arXiv,Dong2014arXiv,LiuBo2014arXiv}%
.

In this Letter, we investigate whether a Weyl point can develop a nontrivial
structure at zero energy and whether there exist any topological property
protecting the developed structure. (i) We first consider a toy model to
examine the possibility for a Weyl point to develop a gapless structure.
Mathematically, the structured Weyl point can be viewed as a bouquet of two
spheres (or wedge of two spheres)~\cite{Hatcher}, which is a new class of
topological state that has not been explored previously. Amazingly, the
zero-energy bouquet is characterized by three distinct topological
invariants: the first Chern number of a surface enclosing the whole bouquet,
the zeroth Chern numbers of the interiors of the two spheres, and the
winding number of a loop enclosing the touching point in the plane of
symmetry. (ii) We further show that the structured Weyl points can be
physically realized in the quasiparticle excitation spectrum of a 3D
spin-orbit-coupled (SOC) fermionic cold-atom superfluid with the
Fulde-Ferrell (FF) ground state. FF superfluids~\cite%
{Zheng2013PRA,FanWu2013PRL,Liu2013PRA,Lin2013NJP,Qu2013NC,Yi2013NC,XJ2013PRA,Chun2013PRL,MGong2014PRB,YGao2014arXiv,YongBKT,YongReveiw}
have been studied recently in SOC degenerate Fermi gases subject to Zeeman
fields, which yield asymmetries of the Fermi surface and induce the FF
Cooper pairing with nonzero total momenta. We obtain a rich phase diagram in
the gapless region of 3D FF superfluids, where not only the featureless Weyl
points~\cite{Gong2011prl,Balents2012PRB,Tumanta2012PRB,
Das2013PRB,Sumanta2013PRA,Yong2014PRL,Yang2014PRL,Balicas2013arXiv,Dong2014arXiv,LiuBo2014arXiv}
but also the structured Weyl points emerge. 
Note that the featureless Weyl points have been well studied in SOC FF superfluids \cite{Yong2014PRL}, 
and here we focus only on the novel structured Weyl points. (iii) We also discuss how the structured Weyl points can be
detected in experiments by measuring spectral densities in photoemission
spectroscopy that has already been utilized in degenerate Fermi gases~\cite%
{Jin08Nature}.

\begin{figure}[t]
\includegraphics[width=3.4in]{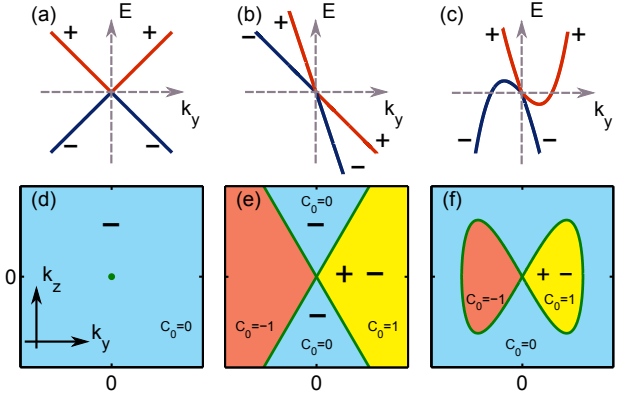}
\caption{The energy dispersions of $H_W$ along $k_{y}$ ($k_x=k_z=0$) for $(%
\protect\alpha,\,\protect\gamma)=(0,0)$ in (a), ($2,0$) in (b), and ($2,1$)
in (c). (d-f) plot the distributions of the negative-energy bands in the $%
k_{x}=0$ plane, corresponding to (a-c) respectively. The states at the green
dots and lines have zero energies. $\pm$ represent the band helicities.}
\label{WeylScheme}
\end{figure}

\emph{Toy model of structured Weyl point}--- Near the Weyl point a Weyl
semimetal or superfluid can be described by Weyl Hamiltonian $%
H_{W}=\pm\sum\nolimits_{i=x,y,z}v_{i}k_{i}\sigma _{i}$, where $\sigma_{i}$
are Pauli matrices and $\pm $ denote the chirality. Clearly, the two bands
disperse linearly and cross only at the Weyl point at the zero energy.
Hereafter we will focus on the positive chirality and choose $v_{i}=1$ for
simplicity. The topological charge of the Weyl point can be characterized by
the first Chern number
\begin{equation}
C_2=\frac{1}{2\pi}\sum_{E_{n}<0}\oint_{\mathcal{S}}{\bm \Omega}_{n}(\bm %
k)\cdot d{\bm S},  \label{ChernEq}
\end{equation}
where the surface $\mathcal{S}$ only encloses the considered Weyl point, and
${\bm \Omega}_{n}(\bm k)=i\langle \nabla_{\bm k} u_{n}(\bm k)|\times
|\nabla_{\bm k} u_{n}(\bm k)\rangle$ is the Berry curvature~\cite{XiaoRMP}
for the $n$-th band with $|u_{n}(\bm k)\rangle$ being its wave function. For
the linearized two-band model, ${\bm \Omega}_{\mp}(\bm k)=\pm{\bm k}/2k^{3}$
and $\mp$ label the eigenvalue of $\sum_{i}k_{i}\sigma _{i}/k$, i.e., the
helicity, as depicted in Fig.~\ref{WeylScheme}(a). This yields $C_2=1$ for
Weyl points with positive chirality.

To generate a structured Weyl point, we add two new momentum dependent terms
in $H_{W}$ such that
\begin{equation}
H_{W}=-\alpha k_{y}\sigma _{0}+(k_{y}+\gamma k_{y}^{3})\sigma
_{y}+k_{x}\sigma _{x}+k_{z}\sigma _{z},  \label{Ham}
\end{equation}%
with $\alpha ,\gamma >0$. The first term that breaks the chiral symmetry
along $y$ is $-\alpha k_{y}\sigma _{0}$, which does not change the
eigenstates and their helicities, leaving the Berry curvatures invariant.
However, this term does change the band energies $E_{\mp }$. With increasing
$\alpha $, the two bands along the ${k}_{y}$ axis rotate clockwise. When $%
\alpha >1$, the particle and hole portions are inverted for the band with
positive velocity, as shown in Fig.~\ref{WeylScheme}(b). Since the helicity $%
+$ and $-$ bands are both occupied at the same momentum and their Berry
curvatures are opposite, their contributions to $C_{2}$ become vanishing. In
Fig.~\ref{WeylScheme}(d) and~\ref{WeylScheme}(e), we plot the distribution
of the occupied-band helicities in the $k_{x}=0$ plane. For $\alpha >1$,
both $+$ and $-$ bands are occupied in the yellow region when $k_{y}>\sqrt{%
(k_{x}^{2}+k_{z}^{2})/(\alpha ^{2}-1)}$, whereas no band is occupied in the
mirror reflected red region. This is in sharp contrast to the case for $%
\alpha <1$, where only the $-$ band is occupied beyond the Weyl point.
Evidently, $C_{2}$ is suppressed and not quantized for $\alpha >1$, which is
anomalous.

To restore a quantized $C_{2}$, we add the cubic term to regularize $H_{W}$.
It follows that the band energies become $E_{\pm }=-\alpha k_{y}\pm E_{0}$
with $E_{0}=\sqrt{(k_{y}+\gamma k_{y}^{3})^{2}+k_{x}^{2}+k_{z}^{2}}$, and
that the Berry curvatures read $\Omega _{{\mp },x}=\pm k_{x}(1+3\gamma
k_{y}^{2})/2E_{0}^{3}$, $\Omega _{{\mp },y}=\pm k_{y}(1+\gamma
k_{y}^{2})/2E_{0}^{3}$, and $\Omega _{{\mp },z}=\pm k_{z}(1+3\gamma
k_{y}^{2})/2E_{0}^{3}$. At large $k_{y}$ the cubic term changes faster than
the linear terms, resulting in a finite zero-energy surface, as sketched in
Fig.~\ref{WeylScheme}(c) and (f), instead of an infinite cone structure.
This suggests that a Weyl point can develop a pair of non-degenerate gapless
spheres, i.e., a \textit{bouquet of two spheres}. We further obtain $C_{2}=1$%
, as long as the surface $\mathcal{S}$ in Eq.~(\ref{ChernEq}) encloses the
whole bouquet.

Apart from the first Chern number $C_2$, intriguingly, there exist two
additional topological invariants characterizing the whole bouquet. In the $%
k_y=0$ plane, there is a chiral symmetry $\sigma_y H_W\sigma_y=-H_W$ and
thus in general $H_W$ can be transformed to an off-diagonal block form $%
\{\{0,h(\mathbf{k})\},\{h(\mathbf{k})^\dagger,0\}\}$ with a winding number
defined as
\begin{equation}
C_1=-\frac{1}{2\pi i}\oint_\mathcal{L}\text{Tr}[h(\mathbf{k})^{-1}\mathbf{d}%
h(\mathbf{k})]\in\mathbb{Z},  \label{EqWN}
\end{equation}
for a loop $\mathcal{L}$ around the band crossing node~\cite%
{Jeffrey,Fan2014PRB,Prodan2014PRB}. Direct calculation yields $C_1=1$ in our
case, and the corresponding Berry phase is $\pi$. Moreover, the whole
bouquet divides the momentum space into three regions with full energy gaps.
Any point $\mathcal{P}$ in these regions is characterized by its zeroth
Chern number~\cite{Jeffrey}, which reads
\begin{eqnarray}
C_0=\frac{1}{2}\left[\sum_{E_{n}<0} \langle u_{n}(\mathcal{P})|u_{n}(%
\mathcal{P})\rangle- \sum_{E_{n}>0} \langle u_{n}(\mathcal{P})|u_{n}(%
\mathcal{P})\rangle\right]. \quad
\end{eqnarray}
$C_0$ amounts to half of the number difference between the occupied and
unoccupied bands since the 0D manifold is featureless in momentum. We find
that in the interior $C_0=1$ ($C_0=-1$) for $k_y>0$ ($k_y<0$) whereas in the
exterior $C_0=0$. Therefore, the whole bouquet can be characterized by $%
(C_0,\,C_1,\,C_2)$.

\emph{Realization in 3D FF superfluids}--- The toy model~(\ref{Ham}) may be
applied to describe the band structures of solid-state materials or the
quasiparticle spectra in superfluids or superconductors. Here we explore the
latter possibility in realizing the structured Weyl points in a 3D SOC
degenerate Fermi gas subject to Zeeman fields, where the dominant ground
state phase is the FF superfluid~\cite%
{Zheng2013PRA,FanWu2013PRL,Liu2013PRA,Lin2013NJP,YongReveiw}.

\begin{figure}[t!]
\includegraphics[width=3.4in]{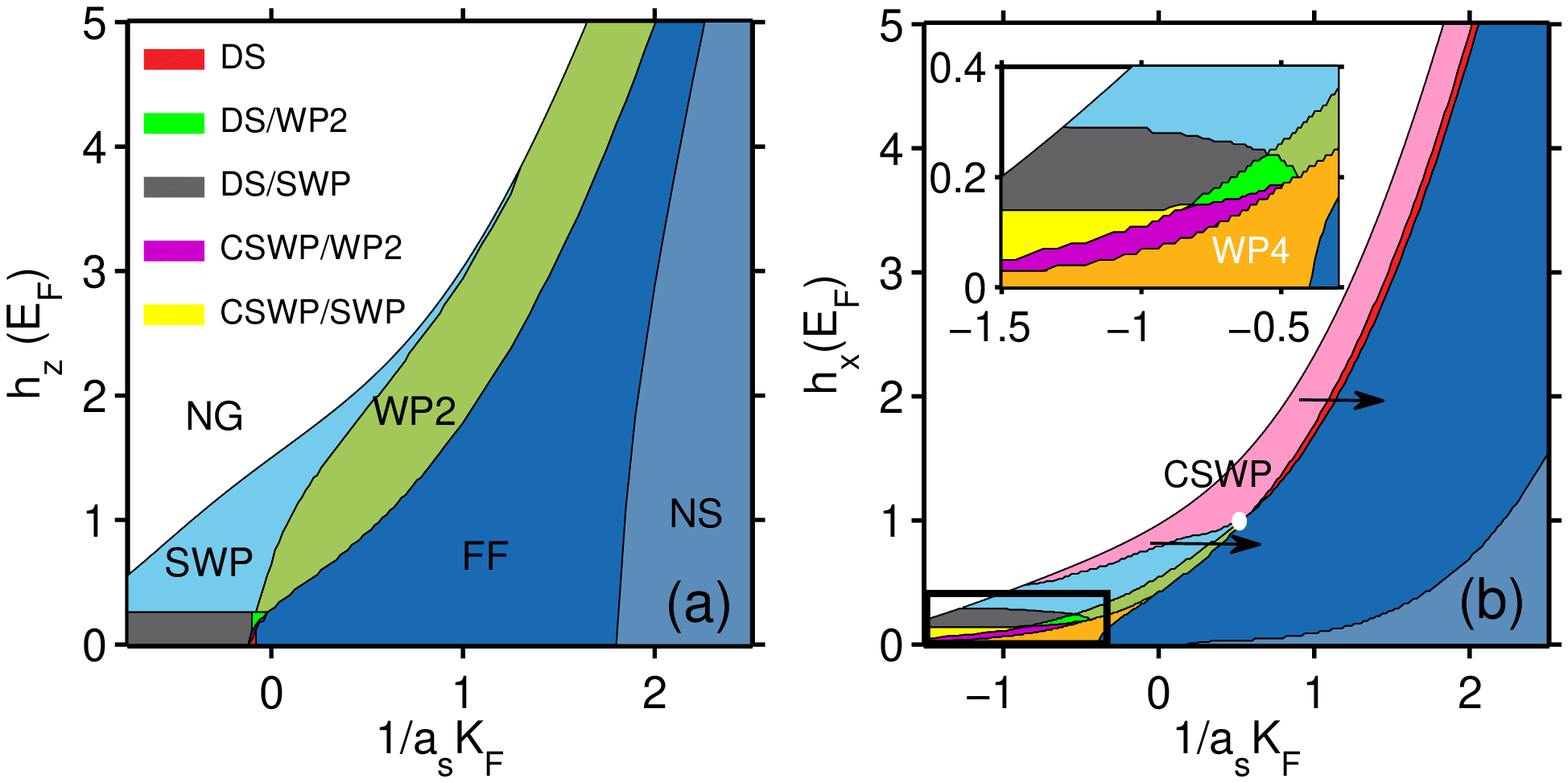} %
\includegraphics[width=3.4in]{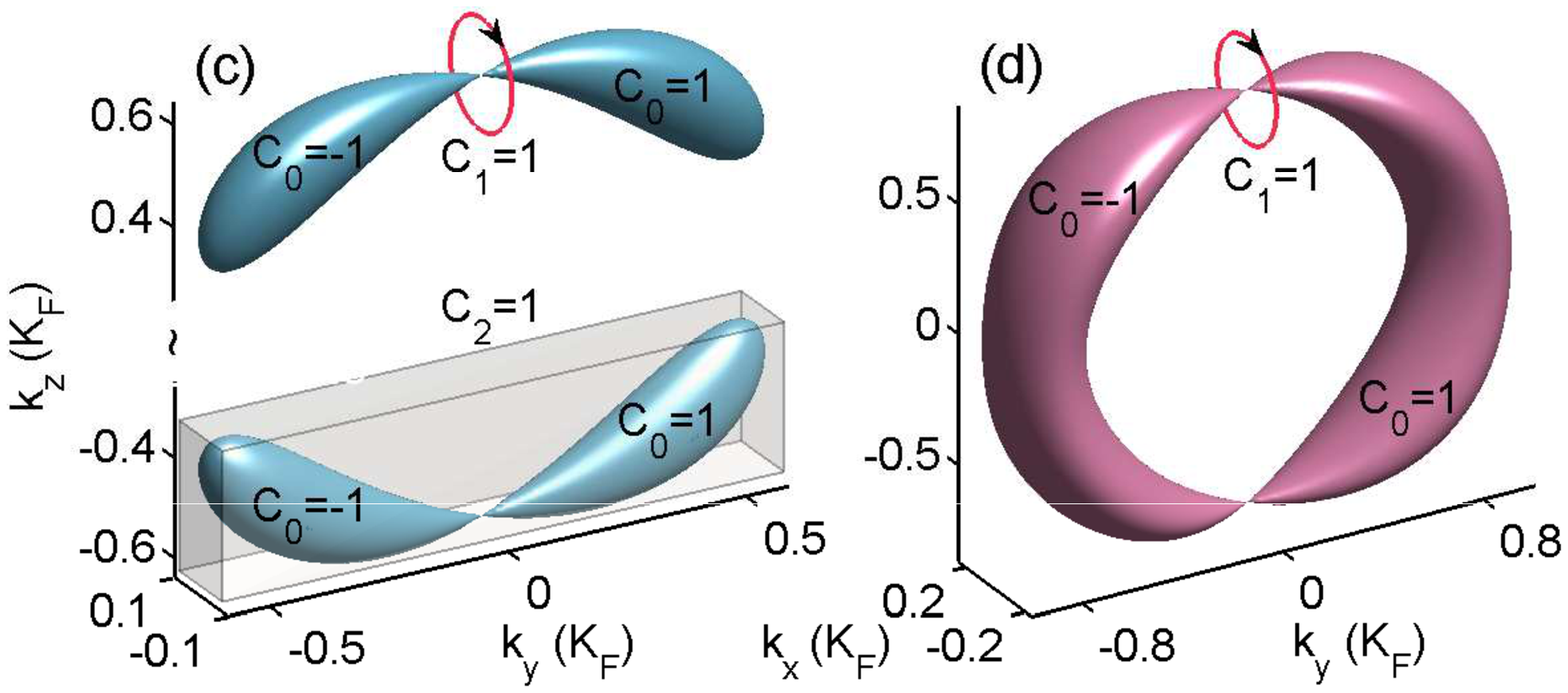} %
\includegraphics[width=3.4in]{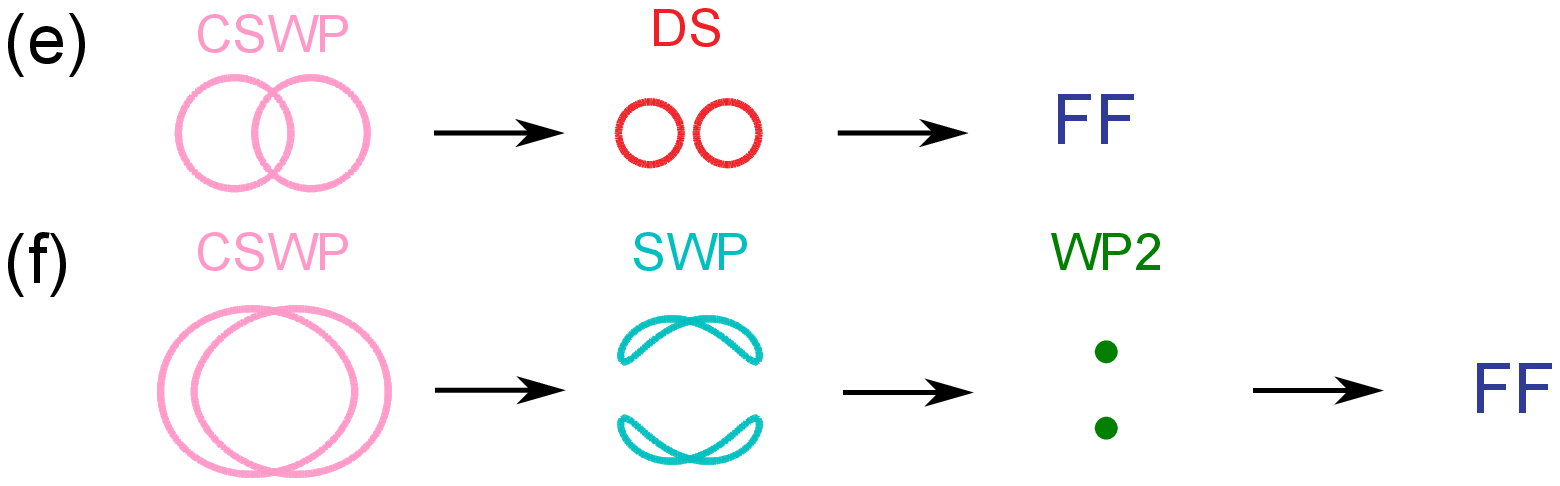} %
\includegraphics[width=3.4in]{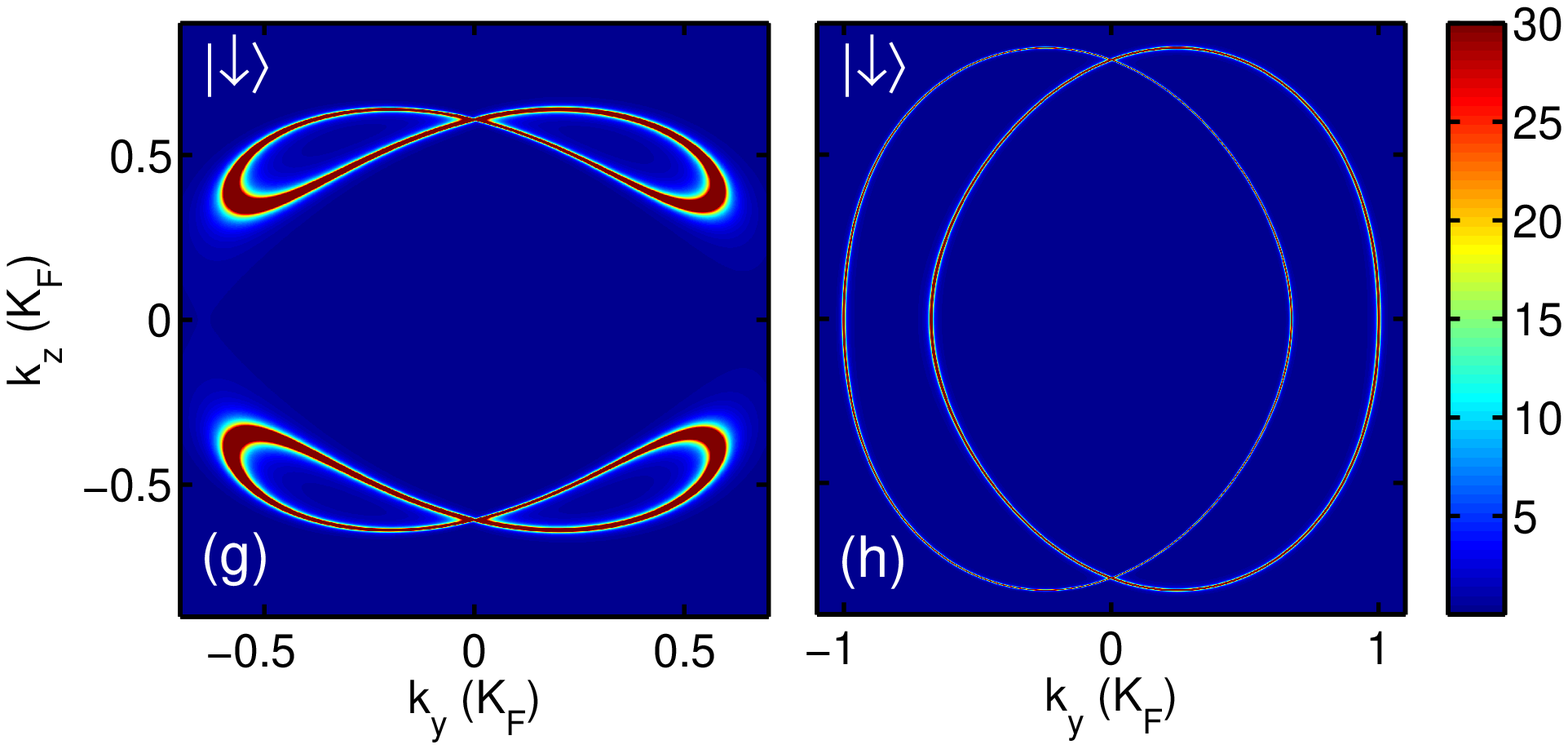}
\caption{Phase diagrams across the BCS-BEC crossover for $h_{x}=0.5E_{F}$ in
(a) and for $h_{z}=0.4E_{F}$ in (b). The inset~\protect\cite{touching4}
enlarges the black rectangular area. Zero-energy contours for a pair of SWPs
in (c) and for a CSWP in (d). (e) and (f) sketch the evolutions of the
zero-energy contours in the $k_{x}=0$ plane along the arrows above and below
the multicritical point (the white point). (g) and (h) show the spin-down
quasiparticle spectral densities in the $k_{x}=0$ plane for (c) and (d). The
Fermi energy $E_{F}$ is $\hbar^{2}K_{F}^{2}/2m$ with $K_{F}=(3\protect\pi%
^{2}n)^{1/3}$. WP2: two Weyl points; WP4: four Weyl points; SWP: structured
Weyl point; CSWP: connected structured Weyl points; DS: disconnected
spheres; NS: normal superfluids with $Q_{y}<10^{-4}K_{F}$; NG: normal gases.
Here $\protect\lambda K_{F}=E_{F}$. }
\label{phase}
\end{figure}

We consider a 3D Fermi gas with a Rashba SOC and an attractive $s$-wave
contact interaction. Based on a standard mean-field approximation~\cite%
{Lin2013NJP}, the thermodynamical potential is $\Omega =|\Delta
_{0}|^{2}/U+\sum\nolimits_{\bm{k}}\left( \varepsilon _{{\bm k}-{\bm Q}%
/2}-\mu \right) -\sum\nolimits_{\bm{k},\sigma }(2\beta )^{-1}\ln
(1+e^{-\beta E_{\bm{k}\sigma }})$. Here $U$ is the interaction strength \cite%
{StringariRMP}, $\beta =1/k_{B}T$ is the inverse of temperature, $%
\varepsilon _{\bm k}={\hbar ^{2}\mathbf{k}^{2}}/{2m}$ is the kinetic energy,
$\mu $ is the chemical potential, $\bm Q$ is the finite momentum of Cooper
pairs with an FF order strength $\Delta _{0}$, and $E_{\mathbf{k}\sigma }$
are the eigenenergies of Bogoliubov-de Gennes (BdG) Hamiltonian
\begin{eqnarray}
H_{BdG} &=&\left[ \varepsilon _{\bm k}-\bar{\mu}+\lambda ({\bm\sigma }\times
{\bm k})\cdot \hat{z}\right] \tau _{z}+\Delta _{0}\tau _{x}  \notag \\
&+&\bar{h}_{x}\sigma _{x}+h_{z}\sigma _{z}+{\hbar ^{2}k_{y}Q_{y}}/{2m},
\label{BdG}
\end{eqnarray}%
with $\bar{\mu}=\mu -Q_{y}^{2}\hbar ^{2}/8m$ and $\bar{h}_{x}=h_{x}+\lambda
Q_{y}/2$. In Eq.~(\ref{BdG}), the Pauli matrices $\bm\sigma $ and $\bm\tau $
act on the spin and the Nambu spaces, $h_{x}$ and $h_{z}$ are the in-plane
and the out-of-plane Zeeman fields, $\lambda $ is the SOC strength, and the
basis is chosen as $(c_{{\bm k}+{\bm Q}/2\uparrow },c_{{\bm k}+{\bm Q}%
/2\downarrow },c_{-{\bm k}+{\bm Q}/2\downarrow }^{\dagger },-c_{-{\bm k}+{%
\bm
Q}/2\uparrow }^{\dagger })^{T}$ with $\bm{Q}=Q_{y}\hat{y}$ ~\cite%
{Zheng2013PRA,Yong2014PRA}. To obtain the self-consistent mean-field
solutions of $\Delta _{0}$, $Q_{y}$, and $\mu $, we solve the nonlinear
saddle point equations $\partial \Omega /\partial \Delta _{0}=0$ and $%
\partial \Omega /\partial Q_{y}=0$, and the atom number equation $\partial
\Omega /\partial \mu =-n$ with a conserved total atom density $n$.

To shed light on how the Weyl nodes emerge and evolve in the BdG spectrum,
we first analyze the symmetries of the physical system beyond the intrinsic
particle-hole symmetry ($\Xi =\sigma _{y}\tau _{y}K$ with $K$ the complex
conjugation). When $h_{x,z}=0$, the time-reversal symmetry ($\Theta =\sigma
_{y}K$) is present, and the mirror symmetries ($\mathcal{M}_{\nu }=-i\sigma
_{\nu }$,\thinspace $\nu =x,y$) are also unbroken by the SOC or the pairing,
i.e., $\mathcal{M}_{\nu }^{-1}H_{BdG}\mathcal{M}_{\nu }=H_{BdG}(-k_{\nu })$.
When the Zeeman fields $h_{x,z}$ are turned on, the time-reversal and the
two mirror symmetries are broken explicitly, since there are terms in the
second line of Eq.~(\ref{BdG}) which are odd under their individual
operations. Yet, these terms are still even under the product operations of $%
\Theta $ and $\mathcal{M}_{y}$. Therefore, independent of the presence of
Zeeman fields, $\Pi =\Xi \Theta \mathcal{M}_{y}=i\sigma _{y}\tau _{y}$ is an
effective symmetry of the system, i.e., $\Pi ^{-1}H_{BdG}\Pi
=-H_{BdG}(-k_{y})$. Similarly, without $h_{x}$ fields (thus $Q_{y}=0$), $%
\mathcal{M}=\Theta \mathcal{M}_{x}$ is an extra symmetry, because of $%
\mathcal{M}^{-1}H_{BdG}\mathcal{M}=H_{BdG}(-k_{y},\pm k_{z})$ (considering
the intrinsic symmetry $H_{BdG}=H_{BdG}(-k_{z})$).

$h_{x}$ Zeeman field is critical for the presence of structured Weyl points.
When $h_{x}=0$, the two symmetries, $\Pi$ and $\mathcal{M}$, dictate that
any zero-energy state of a \textit{dispersed} band must have a zero-energy
partner state at the same momentum. This indicates that a doubly degenerate
Weyl node can exist whereas a non-degenerate state cannot stay at the zero
energy. When $h_{x}\neq 0$, the $\mathcal{M}$-symmetry is broken, and the
double degeneracy for gapless states is only dictated in the $k_y=0$ plane.
Indeed, we find that in the phase diagram two or four Weyl nodes can appear
in this plane, and each can evolve into a band crossing node with a
non-degenerate gapless surface structure developed on each side of the
plane. For Eq.~(\ref{BdG}), a pair of band crossing nodes appears when $%
h_{z}^2>\bar\mu^{2}+\Delta_{0}^{2}-\bar{h}_{x}^{2}$, and two pairs appear
when $\Delta_{0}^{2}-\bar{h}_{x}^{2}<h_{z}^2<\bar\mu^{2}+\Delta_{0}^{2}-\bar{%
h}_{x}^{2}$ and $\bar\mu>0$. We map out the zero-temperature phase diagram
across the BCS-BEC crossover as a function of $h_{z}$ in Fig.~\ref{phase}(a)
and a function of $h_{x}$ in Fig.~\ref{phase}(b). The FF superfluids of
finite-momentum Cooper pairs are dominant except in a region deep into the
BEC side. In sharp contrast to the superfluids without the $h_x$ field where
only Weyl points can exist, these superfluids can also possess structured
Weyl points, as plotted in Fig.~\ref{phase}(c).

As discussed in the toy model, a structured Weyl node can be characterized
by three independent topological invariants $(C_{0},\,C_{1},\,C_{2})$. We
find that $C_{2}=-1$ ($C_{2}=1$) for a closed surface enclosing the whole
bouquet at $k_{z}>0$ ($k_{z}<0$). This reflects that the gapless bouquet
originates from a Weyl node. Furthermore, in the $k_{y}=0$ plane, there is a
chiral symmetry $\Pi^{-1}H_{BdG}\Pi =-H_{BdG}$, and based on Eq.(\ref{EqWN})
the winding number $C_{1}=1$ and the corresponding Berry phase is $\pi $.
Intriguingly, for each bouquet, the non-degenerate gapless surface separates
the momentum space into three topologically distinct regions with different $%
C_{0}$ indices. In the exterior $C_{0}=0$ whereas in the interior $C_{0}=1$ (%
$C_{0}=-1$) on the $k_{y}>0$ ($k_{y}<0$) sides. This result is consistent
with the $\Pi $ symmetry that relates the positive and negative energy bands
at opposite $k_{y}$. Evidently, a whole bouquet has topological invariants $%
(\pm 1,\,\pm 1,\,\pm 1)$. As we have observed, all these unique features of
the structured Weyl node have been exhibited in our toy model~(\ref{Ham}).

As the interaction strength or the Zeeman field strength changes, the
gapless spheres of different bouquets can even connect and merge together as
long as their $C_0$ indices are the same. An example of such a connected
bouquet (i.e. CSWP phase) is shown in Fig.~\ref{phase}(d). The two bouquets
have opposite $C_2$ indices before their merging, and hence any surface
enclosing the whole connected bouquet has $C_2=0$ after merging. Yet, the
winding number $C_1$ of any loop enclosing either band crossing node in the $%
k_y=0$ plane as well as the zeroth Chern number $C_0$ are still intact.
Thus, a connected bouquet has topological invariants $(\pm 1,\,\pm 1,\,0)$.

There exists a multicritical point (the white point) in Fig.~\ref{phase}(b)
where five distinct phases intersect and the gap closes at $\mathbf{k}=0$.
The increasing of $1/a_{s}K_{F}$ for a fixed $h_{x}$ above and below the
multicritical point represent two different processes of destroying the
phase with a connected bouquet. For the larger $h_{x}$, as the interaction
increases, the two band crossing nodes of the connected bouquet move toward
each other along $\hat{k}_{z}$ and then pair-annihilate, leading to a
gapless FF phase with two disconnected spheres (i.e. DS phase in Fig.~\ref%
{phase}) located at two sides of the $k_{y}=0$ plane. As the interaction
further increases, the spheres shrink and eventually disappear, yielding a
gapped FF superfluid. This process is sketched in Fig.~\ref{phase}(e). The
other process for the smaller $h_{x}$ is sketched in Fig.~\ref{phase}(f). As
the interaction increases, the connected bouquet is first broken down into a
pair of disconnected bouquets, then each shrinks into a Weyl node, and
eventually the two Weyl nodes pari-annihilate at ${\bm k}=0$ producing a
gapped FF superfluid. Intriguingly, each gapless phase has at least one
topological invariant.

\begin{figure}[t]
\includegraphics[width=3.4in]{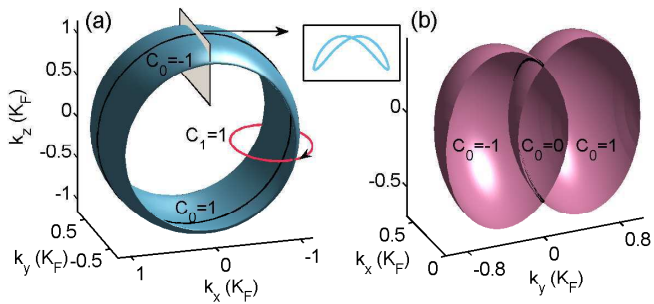}
\caption{(a) Zero-energy contours of a structured Weyl ring (SWR). The inset
shows the cross section of SWR. (b) A closed SWR (CSWR). Only $k_x>0$ part
is plotted; the $k_x<0$ part is symmetric to the $k_x>0$ part with respect
to the $k_x=0$ plane. Here $\protect\lambda K_{F}=E_{F}$. }
\label{SWeylRing}
\end{figure}

\emph{Structured Weyl ring.}--- Given the existence of structured Weyl
points, one may wonder whether there exist structured Weyl rings in 3D. We
can construct a simple model to describe such a ring: $H_{WR}=-\alpha
k_{y}\sigma _{0}+(k_{y}+\gamma k_{y}^{3})\sigma _{y}+(\mathbf{k}%
^{2}-m^{2})\sigma _{z}$ with nonzero $m$. Analogous to structured Weyl
points, when $\alpha >1$ and $\gamma >0$, the zero-energy contour is a
structured Weyl ring, i.e., a bouquet of two circles travels along a loop as
shown in Fig.~\ref{SWeylRing}(a). Although the first Chern number $C_{2}$
for a surface enclosing the whole ring is zero, the Berry phase along the
loop trajectory enclosing the ring is $\gamma =\pi $~\cite{Qi2008PRB} and
the zeroth Chern number $C_{0}$ of the gapped interiors are quantized to $%
\pm 1$. Such a structured Weyl ring can indeed be realized in the
quasiparticle spectrum of SOC FF superfluids, with equal Rashba and
Dresselhaus SOC, i.e., the model~(\ref{BdG}) with the Rashba SOC replaced by
$H_{\text{SOC}}(\hat{\mathbf{p}})=\lambda \hat{p_{y}}\sigma _{x}$~\cite%
{Melo2011PRL}. Here $\Pi $ symmetry and rotational symmetry with respect to $%
k_{y}$ [$H_{BdG}=H_{BdG}(k_{x}^{2}+k_{z}^{2},k_{y})$] dictate that crossing
rings can exist in the $k_{y}=0$ plane. Indeed, in this plane for FF
superfluids with a $h_{x}$ field, a ring appears when $h_{z}^{2}>\bar{\mu}%
^{2}+\Delta _{0}^{2}-\bar{h}_{x}^{2}$, and two rings appear when $\Delta
_{0}^{2}-\bar{h}_{x}^{2}<h_{z}^{2}<\bar{\mu}^{2}+\Delta _{0}^{2}-\bar{h}%
_{x}^{2}$ and $\bar{\mu}>0$. Similar to the case for structured Weyl points,
$\mathcal{M}$-symmetry is broken by $h_{x}$ fields so that the Weyl ring can
develop structures on both sides of the $k_{y}=0$ plane, as shown in Fig.~%
\ref{SWeylRing}(a). The developed gapless contours can further bend toward $%
k_{x}=k_{z}=0$ and become closed, as shown in Fig.~\ref{SWeylRing}(b). It
follows that then only $C_{0}$ is responsible for the protection of the
zero-energy structures.

\emph{Experimental observation.}--- To observe the structured Weyl nodes or
rings, we consider the quasiparticle spectral density $A_{\sigma }(\omega ,%
\bm{k})=-\mathrm{Im}G_{\sigma \sigma }(i\omega =\omega +i\delta ,\bm{k})/\pi
$, where $G=(i\omega -H_{BdG})^{-1}$ is the Green function. $A_{\sigma
}(\omega ,\bm{k})$ can be experimentally measured using the spin and
momentum resolved photoemission spectroscopy~\cite{Jin08Nature}. To evaluate
the zero-energy spectral density, we compute $A_{\downarrow }(\omega =0,%
\bm{k})$ in the $k_{x}=0$ plane, which is illustrated in Fig.~\ref{phase}(g)
and (h) for a pair of structured Weyl nodes. The signal of $A_{\uparrow
}(\omega =0,\bm{k})$ is similar, though it has a weaker amplitude due to the
smaller particle density. For a structured Weyl ring, the signal of $%
A_{\sigma }$ in the $k_{x}=0$ plane is similar to that of the structured
Weyl node, but exhibits a ring structure in the $k_{y}=0$ plane instead of
two points for a pair of structured Weyl nodes. We note that a structured
Weyl node or ring also has an interesting surface spectral density~\cite%
{Yang2014PRL}. In experiments, SOC and Zeeman fields have been recently
realized~\cite{Lin2011Nature,Jing2012PRL,
Zwierlen2012PRL,PanJian2012PRL,Qu2013PRA,Spilman2013PRL,
Spilman2013NatRev,Jing2014Review,Jing2015arXiv} in $^{40}$K and $^{6}$Li
fermionic atoms by coupling two hyperfine states via counter-propagating
Raman laser beams. The SOC strength is determined by the wavelength $l_{r}$
of the Raman beams and their relative angle $\theta $ by $\lambda
K_{F}=2(k_{r}\sin (\theta /2)/K_{F})E_{F}$ with $k_{r}=2\pi /l_{r}$. This
strength can be as large as $\lambda K_{F}=2E_{F}$ when we consider a
typical set of parameters: $l_{r}=773$nm and $n=1.8\times 10^{13}$cm$^{-3}$~%
\cite{Jing2012PRL}. The Zeeman fields, depending on the laser beam strength
or the detuning, can also be readily tuned to $E_{F}$. These parameters are
large enough to observe the exotic Weyl phases discovered here. We note that
in a harmonic trap, the quasiparticle spectrum around the center of the trap
can be measured using spatially resolved photoemission spectroscopy assisted
by hollow light technology \cite{GaeblerThesis}.

In summary, we demonstrate that a Weyl point in semimetals or nodal
superfluids can deform into a bouquet of two spheres. We show that such a
structured Weyl point is characterized by three distinct topological
invariants and can be realized in SOC Fermi gases subject to Zeeman fields.
Such nontrivial structured Weyl points have not been discussed previously in
$^{3}$He or other topological superconductors/superfluids. Our discovery
introduces a new class of topological quantum matter and may have great
impacts in both cold-atom and solid-state communities.

\begin{acknowledgments}
\textbf{Acknowledgements}: We would like to thank D. Xiao, Z. Liu, and X. Li
for helpful discussions. Y.X. and C.Z are supported by ARO
(W911NF-12-1-0334) and AFOSR (FA9550-13-1-0045). F.Z. is supported by UT
Dallas research enhancement funds. We also thank Texas Advanced Computing
Center, where our numerical calculations were performed.
\end{acknowledgments}

\begin{acknowledgments}
\textit{Note added}: After we posted this manuscript on arXiv, we noticed
two newly posted preprints \cite{DaiXi2015,Sun2015} on a new type of Weyl
point and its realization in the single particle band dispersion of certain
solid materials (e.g., WTe$_{2}$ and MoTe$_2$), where the Weyl Hamiltonian
around the Weyl point is similar to our toy Hamiltonian Eq.~(\ref{Ham}) with
only linear terms considered.
\end{acknowledgments}


\begin{thebibliography}{99}
\bibitem{touching4} {For the superfluids with four crossing points, there
are five phases in the phase diagram: WP4, CSWP/WP2, CSWP/SWP, DS/WP2, and
DS/SWP. The phase name consists of two parts: the former containing the
inner crossing points while the latter the outer ones. }
\end{thebibliography}
\end{document}